\documentclass[10pt,twocolumn,letterpaper]{article}

\usepackage{cvpr}              

\newcommand{\FLIP}{\protect\reflectbox{F}LIP\xspace}
\newcommand{\flipview}[1]{\text{\protect\reflectbox{F}LIP}_{#1}\xspace}

\newcommand{\figref}[1]{Fig.~\ref{#1}}
\newcommand{\tabref}[1]{Table~\ref{#1}}
\newcommand{\equref}[1]{Eq.~(\ref{#1})}

\newcommand{\secref}[1]{Sec.~\ref{#1}}
\newcommand{\appref}[1]{Appendix~\ref{#1}}

\usepackage{multirow}
\usepackage{colortbl}
\usepackage{xcolor}
\usepackage{graphicx}
\usepackage{wrapfig}
\usepackage{bbm}
\usepackage{cuted}
\definecolor{c1}{HTML}{ffcc99}
\definecolor{c2}{HTML}{fff8ae}

\definecolor{cvprblue}{rgb}{0.21,0.49,0.74}
\usepackage[pagebackref,breaklinks,colorlinks,allcolors=cvprblue]{hyperref}

\def\paperID{} 
\def\confName{CVPR}
\def\confYear{2026}

\title{Depth Peeling for High-Fidelity Gaussian-Enhanced Surfel Rendering}

\author{Keyang Ye \quad\quad Hongzhi Wu \quad\quad Kun Zhou\footnotemark[1]\\
State Key Lab of CAD\&CG, Zhejiang University\\
Hangzhou Research Institute of Holographic and AI Technology\\
{\tt\small yekeyang@zju.edu.cn, hwu@acm.org, kunzhou@acm.org}
}

\begin{document}
\maketitle

\begin{strip}
\centering
\vspace{-45px}
\includegraphics[width=\textwidth]{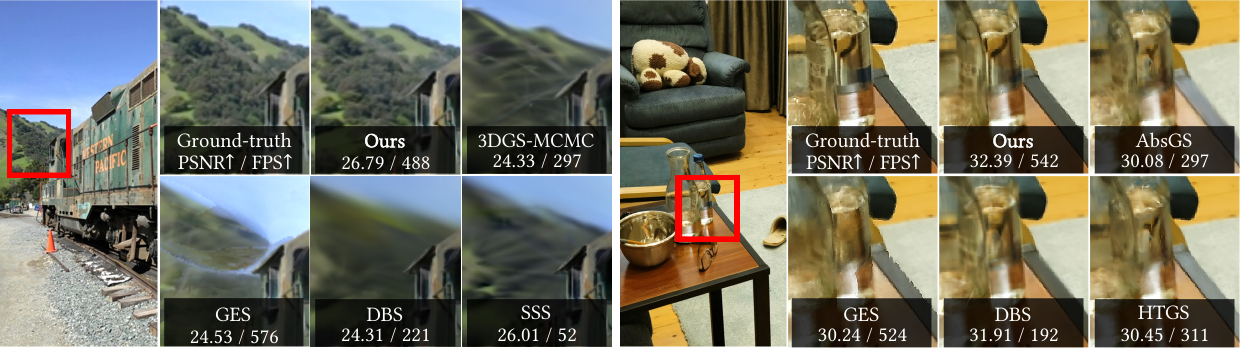}
\captionof{figure}{\small Our novel representation achieves fast, high-fidelity and popping-free rendering, outperforming state-of-the-art techniques: GES~\cite{GES25ye}, SSS~\cite{zhu20253d}, DBS~\cite{Beta25Liu}, AbsGS~\cite{ye2024absgs}, 3DGS-MCMC~\cite{kheradmand20243d} and HTGS~\cite{hahlbohm2025htgs}.}
\label{fig:teaser}
\end{strip}

\footnotetext[1]{Corresponding author}

\begin{abstract}
Novel view synthesis has been significantly advanced by NeRFs and 3D Gaussian Splatting (3DGS), which require ordering volumetric samples or primitives for correct color blending. While the recent Gaussian-Enhanced Surfels (GES) enable high-performance, sort-free rendering, they suffer from aliasing artifacts and suboptimal reconstruction. To address these limitations, we propose DP-GES, a novel representation that combines 2D opaque surfels with semi-transparent boundaries to represent coarse-scale geometry and appearance, and 3D Gaussians surrounding the surfels to supplement fine-scale details. We employ Depth Peeling to achieve accurate per-pixel ordering for surfel rendering, which enables sort-free Gaussian splatting with correct transmittance modulation, effectively eliminating aliasing and popping artifacts while facilitating a fully differentiable joint optimization. Extensive experiments demonstrate that our method achieves superior novel view synthesis quality and compares favorably against state-of-the-art techniques across a wide range of scenes.
\end{abstract}    
\section{Introduction}
\label{sec:intro}

Novel view synthesis from captured images is a fundamental task in computer vision and graphics. The field has seen continuous improvement, evolving from classic light field~\cite{lightfield1996,lumigraph1996} to the recent breakthroughs of Neural Radiance Fields (NeRF)~\cite{nerf} and 3D Gaussian Splatting (3DGS)~\cite{3DGS}. 

Both NeRF and 3DGS represent scenes using volumetric radiance fields and synthesize novel views by blending colors of volumetric samples or primitives in near-to-far order. NeRF relies on volumetric ray marching, which requires computationally expensive neural network inferences at densely sampled points along each ray. This high computational cost significantly limits its applicability in real-time scenarios. 3DGS explicitly sorts Gaussian primitives and employs tile-based sorting to reduce computational overhead. While this approximation improves performance, it introduces view-inconsistent artifacts such as popping, where floaters appear or disappear abruptly due to changes in the sorting order. Subsequent work has explored sort-free rendering of Gaussians to eliminate sorting overhead~\cite{sortfree2024}, but these methods typically suffer from severe occlusion leakage problems.

Recently, Gaussian-Enhanced Surfels (GES)~\cite{GES25ye} and its follow-up~\cite{peng2025gaussian} have emerged as comprehensive solutions for sort-free and view-consistent radiance field rendering. GES employs a hybrid bi-scale representation, modeling coarse geometry and appearance with opaque surfels rendered via a standard z-buffer, while enriching fine details with Gaussians splatted in a sort-free manner and composited against the surfel depth buffer. This approach avoids the popping and leakage artifacts prevalent in prior methods, achieving high performance and visual quality.


Despite its advantages, GES has two key limitations. First, although multisample anti-aliasing (MSAA) mitigates aliasing for surfels, the hard depth test applied to Gaussians truncates their color and weight contributions, resulting in residual aliasing along object boundaries. Second, the joint optimization of surfels and Gaussians is hindered because the geometry of fully opaque surfels is fundamentally non-differentiable, and their colors are only loosely blended with Gaussians, leading to suboptimal reconstruction.

We introduce DP-GES, a novel representation that addresses the limitations of GES by augmenting surfels with semi-transparent boundaries.
Such semi-transparent surfels require sorting for correct blending. To this end, we propose to employ Depth Peeling~\cite{shade1998layered} to achieve accurate per-pixel ordering. This allows Gaussians to be splatted in a sort-free manner while having their weights correctly modulated by the transmittance from corresponding depth layers. Gaussians beyond the last surfel layer or with zero transmittance are culled by the hardware depth test.
Experimentally, peeling only three surfel layers suffices for high-fidelity results, striking a favorable balance between quality and efficiency. Furthermore, the semi-transparent boundaries of surfels enable fully differentiable joint optimization, as the transmittance values through the semi-transparent boundaries link surfels and Gaussians, leading to improved reconstruction.

Extensive experiments demonstrate that our DP-GES achieves high-fidelity novel view synthesis, outperforming or matching state-of-the-art methods across diverse real and synthetic scenes. While introducing only minimal overhead over original GES, our method renders at over 470fps on an RTX 4090 for the Mip-NeRF360 dataset~\cite{mipnerf360}, 2.5$\times$ faster than vanilla 3DGS~\cite{3DGS} and 3$\times$ faster than a recent high-quality splatting approach~\cite{Beta25Liu}.

\section{Related Work}
In this section, we only review previous work most relevant to our method, \ie, 3D Gaussian Splatting and order-independent transparency, while leaving the discussion of NeRF-related work to recent excellent surveys~\cite{gao2022nerf,nerf_survey2025Yun}.

\subsection{3D Gaussian Splatting}
3DGS~\cite{3DGS} achieves real-time and high-quality novel view synthesis by modeling scenes with anisotropic Gaussian primitives, exploiting the efficiency of rasterization for volume rendering. Comprehensive overviews of recent progress can be found in~\cite{wu2024recent, 3dgssurvey,3DGSsurvey-CVM2024}. Here we review previous work on quality improvement to the original 3DGS, particularly on splatting primitives and view-consistent rendering, which are mostly related to this paper. Note that other aspects of improvement, such as densification and pruning strategies~\cite{kheradmand20243d,ye2024absgs} and high-frequency reflection modeling~\cite{deferred3dgs,yang2024spec}, are orthogonal and can be used in conjunction with our pipeline.




\paragraph{Splatting Primitives.}
Recent work explores various kernels to improve geometry~\cite{shen2024solidgs,2dgs2024} or rendering quality~\cite{hamdi2024ges,chen2024linear,held20243d,zhu20253d}.
Deformable beta splatting (DBS)~\cite{Beta25Liu} introduces a deformable radial kernel and replaces the spherical harmonics (SH) color function in 3DGS with spherical Beta, enhancing high-frequency effects and reducing memory footprint.  2D Gaussian Splatting (2DGS)~\cite{2dgs2024} employs 2D Gaussians for accurate surface modeling. SolidGS~\cite{shen2024solidgs} further modifies the 2D Gaussian kernel by enlarging the region with opacity close to 1 for better view consistency. Our surfel kernel is similar to that of SolidGS: except for the boundary, the opacity of our surfel is exactly 1 to facilitate efficient Gaussian culling.
 

Another line of research leverages textured primitives to capture fine-grained texture details, such as HDGS~\cite{song2024hdgs}, Neural Shell Splatting~\cite{zhang2025neural}, Billboard Splatting~\cite{svitov2025billboard}, and Textured Gaussians~\cite{chao2025textured}. These methods improve appearance fidelity with fewer textured primitives, but this comes at a higher cost on per-primitive storage and rendering.

Our DP-GES consists of a small set of 2D surfels and a number of 3D Gaussians, along with a fully differentiable rendering pipeline. This design achieves higher reconstruction quality with lower computational cost. Moreover, these kernel refinements can be directly applied to our Gaussian representation. We incorporate spherical Beta color function from~\cite{Beta25Liu} to improve view-dependent effects.



\paragraph{View-Consistent Rendering.}
3DGS exhibits popping artifacts during camera rotation, since it approximates per-pixel depth ordering with a tile-based global sorting of Gaussian centers for alpha-blending. StopThePop~\cite{stopthepop2024} first proposes a hierarchical sorting strategy to improve sorting precision. DRK~\cite{huang2025deformable} simplifies this strategy by cache sorting. However, approximate sorting cannot fully eliminate popping. Ray-tracing–based methods such as 3DGRT~\cite{moenne20243d} and EVER~\cite{mai2025ever} achieve accurate per-ray ordering but at a much higher computational cost. Inspired by traditional OIT methods, SortFreeGS~\cite{sortfree2024} completely eliminates sorting by blending Gaussians according to the weights computed from depths, but inevitably introduces occlusion leakage. Stochastic Splats~\cite{kheradmand2025stochasticsplats} mitigates popping via stochastic transparency sampling at the cost of temporal noise and reduced rendering efficiency. HTGS~\cite{3dgsht2024} precisely sorts only the first few Gaussians and approximates the rest as a weighted tail. Although this visually reduces popping, it introduces additional overhead due to insertion sort, and prevents early termination. Duplex-GS~\cite{liu2025duplex} introduces cell proxies for local Gaussian grouping, and performs tile-based sorting between cells while aggregating internal Gaussians via weighted sums. Misalignment between cells and Gaussians can degrade quality, and the potential popping caused by cell-level ordering is not thoroughly discussed. 

\paragraph{Comparison with GES.}
GES~\cite{GES25ye} uses a set of fully opaque surfels to build coarse geometry and appearance, while using Gaussians to enrich details. Its surfel is rendered by a standard z-buffer-based pipeline, and Gaussians are accumulated via weighted blending, with surfel depth used for Gaussian culling. This completely eliminates sorting, achieving ultra-fast and popping-free rendering.

Inherited from GES, DP-GES preserves both efficiency and view consistency while addressing its limitations. 
In terms of surfel representation, the opacity in DP-GES smoothly decays from 1 to 0 at the boundaries to mitigate aliasing from depth testing.
For rendering, both surfels and Gaussians are fully differentiable and thus jointly optimized. This leads to improved reconstruction quality.




\subsection{Order-Independent Transparency}
Order-independent transparency (OIT) is a classic problem in real-time rendering. Traditional solutions such as A-Buffer~\cite{abuffer90Haeberli} and K-Buffer~\cite{everitt2001interactive} record multiple ordered fragments per pixel. Depth peeling~\cite{shade1998layered} explicitly renders the scene with multiple passes. For each pass, it ``peels'' away the nearest visible layer by comparing per-pixel depths stored in the previous pass. This process guarantees the correct front-to-back order for alpha compositing. Its memory footprint and rendering cost grow linearly with the number of layers. Weighted blended OIT~\cite{mcguire2013weighted} uses camera-distance-based weights to composite the result. It heavily relies on per-scene weight tuning and is prone to occlusion leakage.

DP-GES combines the advantages of depth peeling and weighted blended OIT: three surfel layers are efficiently peeled to provide accurate depth ordering, while Gaussians are blended in a sort-free manner to accelerate rendering. The peeled depths further guide Gaussian culling, effectively preventing occlusion leakage and enabling fast, robust and view-consistent rendering.



\section{DP-GES Representation}

\begin{figure*}
\includegraphics[width=1.0\linewidth]{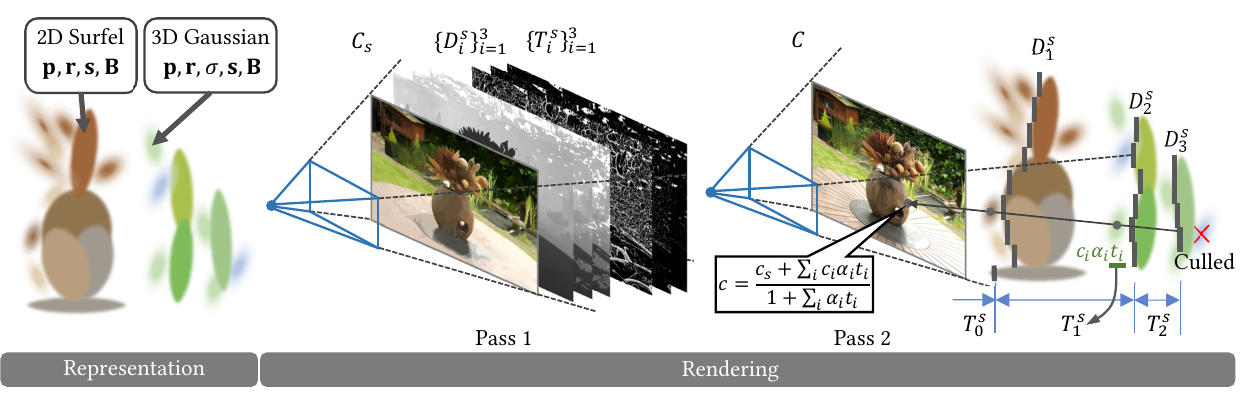}
\caption{Representation and rendering pipeline of DP-GES. The DP-GES representation is composed of a set of 2D opaque surfels with semi-transparent boundaries $\mathcal{S}=\{\mathbf{p}_i, \mathbf{r}_i, \mathbf{s}_i, \mathbf{B}_i\}_{i=1}^N$, and a few 3D Gaussians $\mathcal{G}=\{\mathbf{p}_i, \sigma_i, \mathbf{r}_i, \mathbf{s}_i, \mathbf{B}_i\}_{i=1}^M$ surrounding the surfels, where $\mathbf{p}_i, \sigma_i, \mathbf{r}_i, \mathbf{s}_i, \mathbf{B}_i$ are the position, maximum opacity, rotation, scaling and spherical Beta parameters~\cite{Beta25Liu}, respectively. DP-GES renders in two passes: first, it performs 3-layer depth peeling of surfels using the standard graphics pipeline to obtain surfel colors $C_s$, and per-layer depth $\{D_i^s\}_{i=1}^3$ and transmittance maps $\{T_i^s\}_{i=0}^3$ (with $T_0^s=1$). Then, 3D Gaussians are splatted onto the screen, accumulating their colors and weights by a sort-free additive blending, where each Gaussian queries transmittance value $t_i$ according to its corresponding depth layer, multiplied with its Gaussian weight $\alpha_i$ to get the final weights. \label{fig:pipeline}}
\end{figure*}

Our representation contains two types of primitives: a set of 2D opaque surfels with semi-transparent boundaries to represent the coarse-scale geometry and appearance, and a set of 3D Gaussians surrounding the surfels to supplement fine-scale appearance details. It is an improved variant of GES~\cite{GES25ye}. Below we focus on key differences between the two representations.

\paragraph{Surfel Representation.} Our 2D surfel set $\mathcal{S}$ is defined as $\mathcal{S}=\{\mathbf{p}_i, \mathbf{r}_i, \mathbf{s}_i, \mathbf{B}_i\}_{i=1}^N$, where $\mathbf{p}_i\in\mathbb{R}^3$ is the surfel center position, $\mathbf{r}_i\in\mathbb{R}^4$ is the rotation quaternion, $\mathbf{s}_i\in\mathbb{R}^2$ is the anisotropic scaling, and $\mathbf{B}_i$ is the parameters of the Spherical Beta functions~\cite{Beta25Liu} (SB), representing view-dependent colors. The color of the whole surfel is the same as the color in the direction from the surfel center to the camera position. More details of spherical Beta can be found in~\cite{Beta25Liu}.

\begin{figure}
\includegraphics[width=1.0\linewidth]{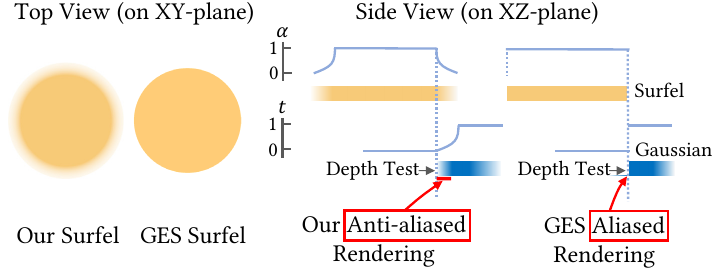}
\caption{Comparison of our surfel representation and the surfel in GES~\cite{GES25ye}. Left: the shape of our 2D opaque surfel with semi-transparent boundary and GES's fully opaque surfel on top view. Right: the alpha $\alpha$ and transmittance $t$ distribution on side view. Although GES employs MSAA, it only smooths the surfel rendering. Once Gaussians are added, the hard depth test acts as if transmittance were abruptly clamped from 1 to 0, reintroducing aliasing. In contrast, DP-GES performs depth testing after Gaussian weights have smoothly decayed through transmittance, thereby reducing aliasing artifacts.\label{fig:surfel_illus}}
\end{figure}


A surfel in DP-GES has a gradually decreasing opacity within a thin ring-shaped region along the boundaries, as shown in~\figref{fig:surfel_illus}. For a surfel defined as a circular disc on the XY-plane in its local coordinates, the opacity for a point $(x,y)$ on the surfel is now defined as: 

$\alpha_i(x,y)=\min(1, wG(x,y))$, with
$G(x,y)=\exp(-\frac{x^2+y^2}{2})$.
Here $w$ is the opacity modulation parameter, a fixed constant shared by all surfels. When $w<1$, the surfel becomes fully translucent, while when $w=255$, it is completely opaque, identical to the original GES surfel. We use a fixed $w=30$ in our final representation, which keeps the central region opaque while preserving a thin semi-transparent boundary ring. 


\paragraph{Gaussian Representation.} The 3D Gaussians $\mathcal{G}$ in DP-GES are defined as $\mathcal{G}=\{\mathbf{p}_i, \sigma_i, \mathbf{r}_i, \mathbf{s}_i, \mathbf{B}_i\}_{i=1}^M$. The only difference from the original GES is that we replace the SH color function with spherical Beta.

\section{DP-GES Rendering}
\label{sec:rendering}
We use the following equation to render an image $C$ as a weighted combination of the colors from surfels and Gaussians, similar to~\cite{GES25ye}: $C = \frac{C_s+C_G}{W_s+W_G}.$
Here $C_S$ and $C_G$ are the colors rendered from surfels and Gaussians, respectively, and $W_S$ and $W_G$ are the corresponding sums of weights. 

As shown in~\figref{fig:pipeline}, the rendering of DP-GES consists of two main stages. In the first stage, we rasterize surfels through a standard graphics pipeline, extracting the nearest 3 layers of depth maps $\{D_i^s\}_{i=1}^3$, color maps $\{C_i^s\}_{i=1}^3$, and opacity maps $\{A_i^s\}_{i=1}^3$ from the surfels by performing z-buffer–based depth peeling. We use 3 layers because, as observed in our experiments, it helps reduce background color leakage in surfel rendering while balancing efficiency and robustness. A discussion is detailed in~\secref{sec:ablation}.

The 3 layers are then alpha-blended, following the same compositing scheme as in 3DGS~\cite{3DGS}, to obtain the final surfel-rendered color $C_s$ and the transmittance maps $\{T_i^s\}_{i=0}^3$ as
\begin{equation}
    C_s=\sum_{i=1}^3A_i^sT_{i-1}^sC_i^s+T_3^sC_b,
\label{equ:s_render}
\end{equation}
\begin{equation}
T_i^s =
\begin{cases}
1, & \text{if } i=0, \\
\prod_{j=1}^{i}(1-A_j^s), & \text{otherwise}.
\end{cases}
\end{equation}
Here $C_b$ is the background color. Therefore, $W_s=\sum_{i=1}^3A_i^sT_{i-1}^s+T_3^s=1$ is the sum of surfel color weights, which is a constant.

In the second stage, we splat the 3D Gaussians onto the screen without sorting, and accumulate their colors and weights for each pixel. 
The accumulated Gaussian color $C_G$ and weight $W_G$ for a pixel $\hat{\mathbf{x}}$ are:
\begin{equation}
    \mathbf{C}_G(\hat{\mathbf{x}})=\sum_{i=1}^K\mathbbm{1}_{dt}(\hat{\mathbf{x}})\mathbf{c}_i\alpha_i(\hat{\mathbf{x}})t_s(\hat{\mathbf{x}}), 
\label{equ:accu_color}   
\end{equation}
\begin{equation}
    W_G(\hat{\mathbf{x}})=\sum_{i=1}^K\mathbbm{1}_{dt}(\hat{\mathbf{x}})\alpha_i(\hat{\mathbf{x}})t_s(\hat{\mathbf{x}}),
\label{equ:accu_weight}  
\end{equation}
where the indicator function $\mathbbm{1}_{dt}(\hat{\mathbf{x}})$ denotes the depth test; $\alpha_i(\hat{\mathbf{x}})$ is the Gaussian weight; $\mathbf{c}_i$ is the Gaussian color in the current view direction, and $K$ is the number of Gaussians covering the pixel. $t_s(\hat{\mathbf{x}})$ is defined as
\begin{equation}
t_s(\hat{\mathbf{x}}) =
\begin{cases}
T_0^s(\hat{\mathbf{x}}), & \text{if } d_i < D_1^s(\hat{\mathbf{x}}), \\
T_1^s(\hat{\mathbf{x}}), & \text{if } D_1^s(\hat{\mathbf{x}}) < d_i < D_2^s(\hat{\mathbf{x}}), \\
T_2^s(\hat{\mathbf{x}}), & \text{otherwise}.
\end{cases}
\label{equ:trans}
\end{equation}
Each Gaussian determines its corresponding depth interval from $\{D_i^s\}_{i=1}^3$ based on its depth at the center. The transmittance value is queried using~\equref{equ:trans} and multiplied with the Gaussian weight to obtain the final weight. This ensures partially occluded Gaussians fade near surfel boundaries, reducing aliasing artifacts as illustrated in~\figref{fig:surfel_illus}.

During rasterization, depth test is performed: any Gaussian whose depth lies beyond the 3rd peeled layer or whose transmittance weight is zero is discarded for correct occlusion handling. We use an indicator function $\mathbbm{1}_{dt}(\hat{\mathbf{x}})=\mathbbm{1}(d_i<d_s(\hat{\mathbf{x}})+\epsilon_s)$ to denote this per-pixel culling process. Here $d_i$ is the depth of a Gaussian center, as we assume a uniform depth $d_i$ for all pixels covered by a Gaussian. $d_s(\hat{\mathbf{x}})$ is the depth used for culling at a pixel $\hat{\mathbf{x}}$ and $\epsilon_s$ is a tiny depth margin (detailed in~\appref{sec:epsilon}). $d_s$ is set to the depth $D_i^s(\hat{\mathbf{x}})$ of the nearest peeled layer $i$ where $T_i^s(\hat{\mathbf{x}})=0$; if all transmittance values of 3 layers at $\hat{\mathbf{x}}$ remain non-zero, $d_s$ is set to a default value of the 3rd peeled depth $D_3^s(\hat{\mathbf{x}})$.



\section{DP-GES Optimization}
\label{sec:optim}
Given multi-view images of a static scene with camera poses~\cite{sfm2016}, we aim to optimize 2D surfels and 3D Gaussians that faithfully reconstruct the scene for novel-view synthesis. We perform a fully differentiable joint optimization of all parameters after a fast initialization of surfels and 3D Gaussians (detailed in~\appref{sec:training}). Our optimization pipeline is built on PyTorch, with an equivalent OpenGL-based renderer for real-time rendering to fully leverage the standard graphics pipeline (detailed in ~\appref{sec:opengl}).

Our loss function for optimization is $\mathcal{L}=\mathcal{L}_{rgb}+\lambda_1\mathcal{L}_s+\lambda_2\mathcal{L}_{scale}+\lambda_3\mathcal{L}_t$. We use $\lambda_1=0.01, \lambda_3=0.08$ for all the scenes. $\lambda_2$ is set to $5\times10^{-5}$ for all unbounded scenes and $1\times10^{-5}$ for all bounded scenes.

Specifically, $\mathcal{L}_{rgb}$ is the same image loss used in 3DGS, computed between the rendered images $C$ and the corresponding ground-truth images $I_{gt}$. In addition, we impose an $\mathcal{L}_1$ constraint $\mathcal{L}_s=\mathcal{L}_1(C_s,I_{gt})$ between the rendered surfel results $C_s$ and the ground-truth images, ensuring that the surfels faithfully fit the coarse appearance and geometry of the scene. To prevent surfels from growing excessively during optimization, which could compromise geometric robustness, we penalize oversized surfels using an exponential function:

$\mathcal{L}_{scale}=\frac{1}{N}\sum_{i=1}^N\exp(\frac{\tilde{\mathbf{s}}_i^X+\tilde{\mathbf{s}}_i^Y}{2})$,

where $\tilde{\mathbf{s}}_i^X$ and $\tilde{\mathbf{s}}_i^Y$ denotes the surfel scaling along the X and Y axes, respectively, after mean normalization across all surfels.

To further mitigate background color leakage in surfel rendering, the final term is defined as:

$    \mathcal{L}_t=\frac{1}{HW}\sum_{\hat{\mathbf{x}}}(1-T_3(\hat{\mathbf{x}}))^2$.

Here $HW$ is the number of pixels. In DP-GES, we peel only 3 surfel layers, and Gaussians with depths beyond the third peeled layer are occluded, which requires the transmittance after the 3rd layer to be zero (\ie, $T_3=0$); otherwise, background color leakage or truncation artifacts may occur. We already improve the probability of $T_3=0$ by retaining a large opaque region of surfels. A natural idea to further enforce this condition would be to add a loss term encouraging $T_3=0$. However, experiments show that this approach actually increases the stacking of semi-transparent regions within a pixel, thereby raising the probability that $T_3\neq0$. 

Instead, we take the opposite strategy by adding $\mathcal{L}_t$. Though this may sound counterintuitive, it is in fact effective: for pixels $\hat{\mathbf{x}}$ where $T_3(\hat{\mathbf{x}})=0$, they must have been covered by an opaque surfel region, which contributes no gradient to its geometry parameters, so the loss has no effect. For pixels $\hat{\mathbf{x}}$ where $T_3(\hat{\mathbf{x}})\neq0$, this loss encourages the overlapping semi-transparent parts of surfels to go apart, making it more likely for this pixel to be covered by opaque regions of other surfels. This essentially drives $T_3$ toward zero. For pixels covered by fewer than 2 surfel layers, we mask out the corresponding loss gradients.

Please refer to~\appref{sec:training} for more details on optimization and initialization strategies.

\begin{figure*}
\includegraphics[width=1.0\linewidth]{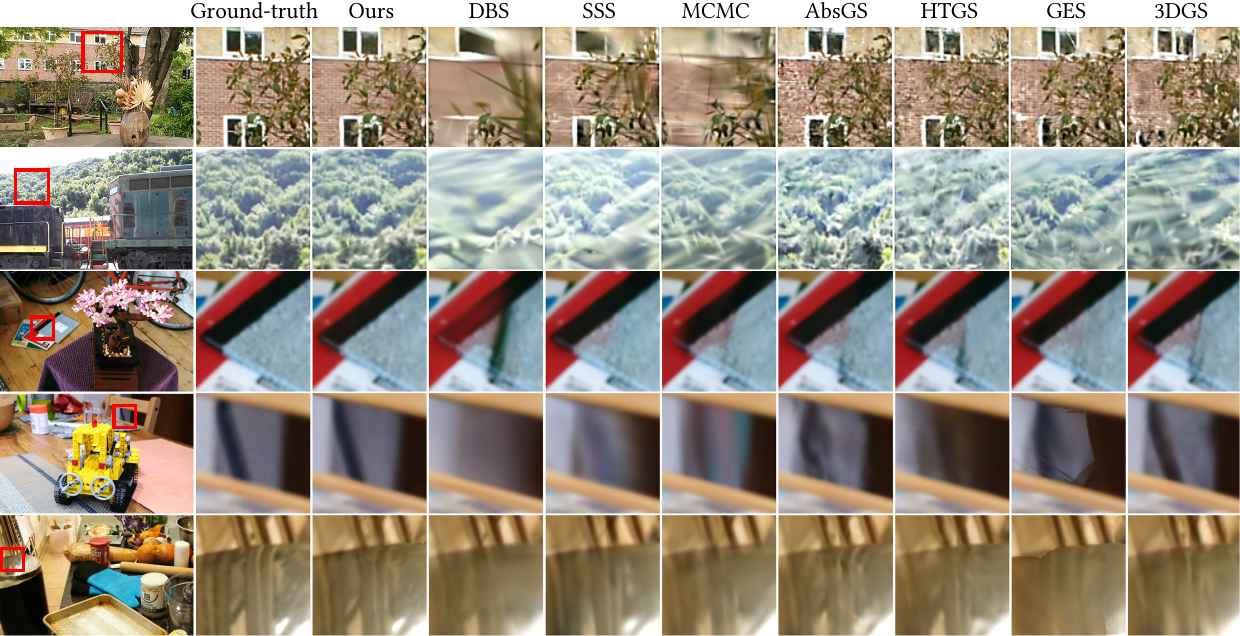}
\caption{Qualitative comparisons on image quality. From top to bottom: \textit{Garden}, \textit{Train}, \textit{Bonsai}, \textit{Kitchen} and \textit{Counter}. Our DP-GES representations reconstruct rich appearance details with joint optimization while maintaining high frame rates with sorting-free rendering.\label{fig:quality_cmp}}
\end{figure*}

\begin{table*}[t]
\caption{Dataset-averaged quantitative evaluation on image quality. Results marked with * are evaluated using our reproduced code.\label{tab:quality_metric}}
\tabcolsep=0.18cm
\footnotesize
\renewcommand\arraystretch{1.1}
\begin{tabular}{l|ccc|ccc|ccc|ccc}
\hline
Datasets      & \multicolumn{3}{c|}{Mip-NeRF360} & \multicolumn{3}{c|}{Deep Blending} & \multicolumn{3}{c|}{Tanks \& Temples} & \multicolumn{3}{c}{NeRF Synthetic} \\
Method/Metric        & SSIM $\uparrow$  & PSNR $\uparrow$  & LPIPS $\downarrow$ & SSIM $\uparrow$       & PSNR $\uparrow$      & LPIPS $\downarrow$     & SSIM $\uparrow$        & PSNR $\uparrow$       & LPIPS $\downarrow$      & SSIM $\uparrow$       & PSNR $\uparrow$      & LPIPS $\downarrow$     \\ \hline
3DGS          & 0.814     & 27.43    & 0.214    & 0.903     & 29.41     & 0.243     & 0.841      & 23.62      & 0.183      & 0.969      & 33.31     & 0.037     \\
AbsGS         & 0.821     & 27.49    & 0.208    & 0.902     & 29.65     & 0.243     & 0.851      & 23.75      & 0.167      & 0.969      & 33.32     & 0.037     \\
MipSplat             & 0.815 & 27.49 & 0.214 & 0.904      & 29.48     & 0.243     & 0.846       & 23.71      & \cellcolor{c1}0.158      & 0.969      & 33.33     & 0.037     \\
3DGS-MCMC     & \cellcolor{c1}0.830     & 27.85    & 0.206    & 0.909     & 29.77     & 0.257     & 0.858      & 24.33      & 0.180      & \cellcolor{c2}0.970      & 33.33     & \cellcolor{c2}0.033     \\
DBS           & \cellcolor{c2}0.827     & \cellcolor{c2}28.10    & 0.210    & \cellcolor{c1}0.912     & \cellcolor{c2}30.25     & 0.250     & 0.860      & 24.52      & 0.166      & \cellcolor{c1}0.972      & \cellcolor{c1}34.34     & \cellcolor{c1}0.030     \\
SSS           & 0.824     & 27.78    & \cellcolor{c2}0.203    & \cellcolor{c2}0.911     & \cellcolor{c2}30.25     & 0.242     & \cellcolor{c2}0.864      & \cellcolor{c1}24.70      & 0.166      & 0.968      & 33.63     & 0.036     \\
GES        & 0.813     & 27.38    & 0.208    & 0.906     & 30.00     & \cellcolor{c2}0.241     & 0.841      & 23.95      & 0.181      & 0.967      & 33.37     & \cellcolor{c2}0.033     \\
SortFreeGS*    & 0.790     & 27.04    & 0.263    & 0.910     & 30.22     & 0.254     & 0.825      & 23.49      & 0.223      & 0.964      & 32.72     & 0.040     \\
StopThePop    & 0.816     & 27.44    & 0.217    & 0.904     & 29.69     & 0.248     & 0.843      & 23.43      & 0.178      & 0.970      & 33.32     & 0.035     \\
HTGS          & 0.820     & 27.04    & 0.235    & 0.908     & 29.88     & 0.301     & 0.849      & 23.17      & 0.206      & 0.969      & 33.73     & 0.036     \\
Ours          & \cellcolor{c2}0.827     & \cellcolor{c1}28.11    & \cellcolor{c1}0.196    & 0.909     & \cellcolor{c1}30.30     & \cellcolor{c1}0.239     & \cellcolor{c1}0.866      & \cellcolor{c2}24.61      & \cellcolor{c2}0.162      & \cellcolor{c2}0.970      & \cellcolor{c2}34.13     & \cellcolor{c1}0.030     \\ \hline
\end{tabular}
\vspace{-10px}
\end{table*}
\section{Results and Evaluation}
We conduct experiments on a workstation with an i7-13700KF CPU, 32GB memory and an NVIDIA RTX 4090 GPU, to demonstrate the effectiveness and efficiency of DP-GES.

\paragraph{Datasets.} We perform tests on the same datasets used in~\cite{3DGS}, including 8 synthetic scenes from NeRF Synthetic Dataset~\cite{nerf}, 9 real-world scenes from Mip-NeRF360 Dataset~\cite{mip_nerf_360}, 2 real scenes from DeepBlending Dataset~\cite{hedman2018deep}, and 2 from Tanks \& Temples Dataset~\cite{knapitsch2017tanks}. The collection covers both synthetic objects with intricate geometries and various texture-rich indoor and outdoor environments, providing a comprehensive benchmark for comparing different approaches. Following~\cite{3DGS}, we use the same pre-downscaled JPG images for training and evaluation on real scenes.

\paragraph{Baselines and Metrics.} We comprehensively compare DP-GES with vanilla \textbf{3DGS}~\cite{3DGS}, 4 methods with state-of-the-art  reconstruction quality:
\textbf{AbsGS}~\cite{ye2024absgs}, 
\textbf{MipSplat}~\cite{mipgs2024}, 
\textbf{3DGS-MCMC}~\cite{kheradmand20243d}, 
\textbf{DBS}~\cite{Beta25Liu} and
\textbf{SSS}~\cite{zhu20253d}, 
4 methods eliminating or reducing popping artifacts:
\textbf{GES}~\cite{GES25ye},
\textbf{SortFreeGS}~\cite{hou2024sortfree}, 
\textbf{StopThePop}~\cite{stopthepop2024},
\textbf{HTGS}~\cite{hahlbohm2025htgs}.
For all MCMC-based methods, including 3DGS-MCMC, DBS and SSS, we set the number of primitives to be equal to the total number of surfels and Gaussians used in our DP-GES for a fair comparison.

We use PSNR, LPIPS, and SSIM metrics to evaluate the rendering quality. We use FPS, computed by the reciprocal of the average rendering time, to evaluate the rendering speed. We also follow StopThePop~~\cite{stopthepop2024} to use $\flipview{1}$ and $\flipview{7}$ to evaluate the short-term and long-term popping artifacts in video results. We employ the same ``flexible stopping" strategy as used in DBS for all baselines. This ensures every method was trained enough iterations to their optimal quality.

Due to space limit, below we focus on our core advantages and key ablation designs. Additional results, including surfel geometry reconstruction, surfel-Gaussian decomposition rendering and detailed ablations, are presented in the supplementary material. We also provide supplementary video results for dynamic visual comparisons.

\subsection{Comparisons}
\begin{figure}[t]
\includegraphics[width=1.0\linewidth]{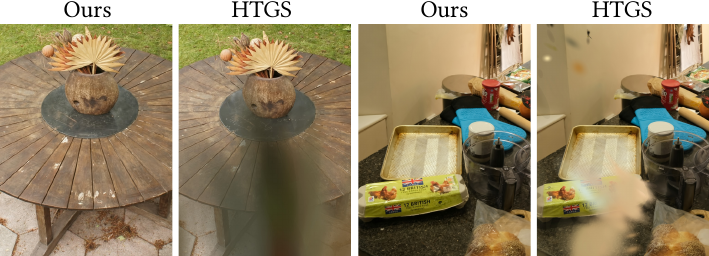}
\caption{Qualitative comparisons with HTGS~\cite{hahlbohm2025htgs}. Floaters with high opacity appear in their results.\label{fig:htgs_cmp}}
\end{figure}

\paragraph{Rendering Quality.} In~\tabref{tab:quality_metric}, DP-GES matches or surpasses all state-of-the-art methods across all datasets, particularly in LPIPS metric. As illustrated in~\figref{fig:quality_cmp}, our method faithfully preserves fine appearance details, especially in distant backgrounds. In \textit{Bonsai}, DP-GES effectively removes the floater artifacts common in other baselines. 
In \textit{Counter}, DP-GES even achieves high-fidelity reflections.
These improvements stem from our representation and fully differentiable rendering, where surfels provide coarse geometry and Gaussians collaboratively refine appearance details.
Among other baselines, MCMC-based methods (DBS, SSS, MCMC) produce blurry results. AbsGS reconstructs details but introduces noise due to its densification strategy. \figref{fig:htgs_cmp} highlights the floater artifacts in HTGS. By sorting only the front 16 Gaussians for alpha blending and approximating the rest as a weighted tail, HTGS avoids popping but amplifies floaters. Floaters often dominate the front depth ranks, pushing deeper Gaussians into the low-weight tail. During optimization, the floaters gradually increase in opacity to compensate for tail errors, making them more noticeable.

\begin{figure}[t]
\includegraphics[width=1.0\linewidth]{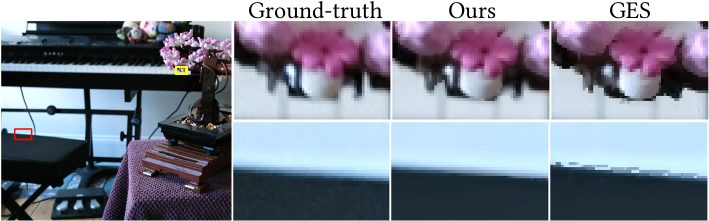}
\caption{Qualitative comparisons with GES~\cite{GES25ye}. GES exhibits noticeable aliasing artifacts due to its depth test.\label{fig:alias_cmp}}
\end{figure}

\begin{figure}[t]
\includegraphics[width=1.0\linewidth]{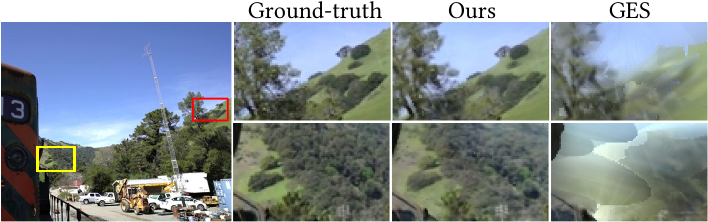}
\caption{Qualitative comparisons with GES~\cite{GES25ye}. GES exhibits poor coordination between surfels and Gaussians, resulting in protruding surfels that occlude fine details.\label{fig:ges_cmp}}
\end{figure}

To more clearly demonstrate our improvements over GES, we present more qualitative comparisons in~\figref{fig:alias_cmp}. Our method produces smoother object boundaries than GES. 
In~\figref{fig:ges_cmp} and \textit{Kitchen} in~\figref{fig:quality_cmp}, some protruding surfels significantly degrade GES’s rendering quality. Surfels in GES are optimized during its surfel optimization stage but cannot be refined during its joint optimization stage. In contrast, surfels and Gaussians are updated together during DP-GES joint optimization, ensuring mutual adaptation and harmonious appearance.

\paragraph{Efficiency.} As reported in~\tabref{tab:fps_metric}, DP-GES achieves over 470 fps at 1080p, which is more than 3$\times$ faster than other high-fidelity baselines such as DBS and SSS. The minor overhead over GES arises mainly from the Gaussian rendering pass, which requires fetching peeled depth and transmittance values from texture buffers, increasing bandwidth usage. In contrast, 3-layer depth peeling adds negligible cost since the number of surfels is much smaller than that of Gaussians (typically less than 10\%), and rasterization within the standard graphics pipeline is highly efficient.


\begin{table}[t]
\centering
\caption{Dataset-averaged quantitative evaluation on frame rate (FPS) at 1080p resolutions, storage (MB), training time (minute) and popping artifacts ($\flipview{1}$ and $\flipview{7}$), averaged on Mip-NeRF360 dataset. Results marked with * are evaluated using our reproduced code. The training time (encompassing all training steps including initialization) is measured for the native image resolutions of training images.\label{tab:fps_metric}}
\footnotesize
\tabcolsep=0.22cm
\renewcommand\arraystretch{1.1}
\begin{tabular}{l|ccc|cc}
\hline
           & FPS$\uparrow$ & MEM$\downarrow$ & Train$\downarrow$ & $\flipview{1}$$\downarrow$  & $\flipview{7}$$\downarrow$  \\ \hline
3DGS       & 185 & 734 & 28    & 0.0250 & 0.0471 \\
AbsGS      & 176 & 746 & 29    & 0.0256 & 0.0473 \\
3DGS-MCMC  & 160 & 348 & \cellcolor{c2}21    & 0.0236 & 0.0472 \\
DBS        & 156 & \cellcolor{c2}165 & \cellcolor{c1}20    & 0.0297 & 0.0771 \\
SSS        & 62  & 351 & 38    & 0.0300 & 0.0716 \\
3D-GES     & \cellcolor{c1}675 & 366 & 43    & \cellcolor{c1}0.0229 & \cellcolor{c1}0.0394 \\
SortFreeGS* & 321 & 506 & 42    & 0.0271 & 0.0625 \\
StopThePop & 167 & 830 & 40    & 0.0247 & 0.0466 \\
HTGS       & 271 & 491 & 23    & 0.0243 & 0.0587 \\
Ours       & \cellcolor{c2}472 & \cellcolor{c1}156 & 40    & \cellcolor{c2}0.0232 & \cellcolor{c2}0.0431 \\ \hline
\end{tabular}
\end{table}

\begin{figure}
\includegraphics[width=1.0\linewidth]{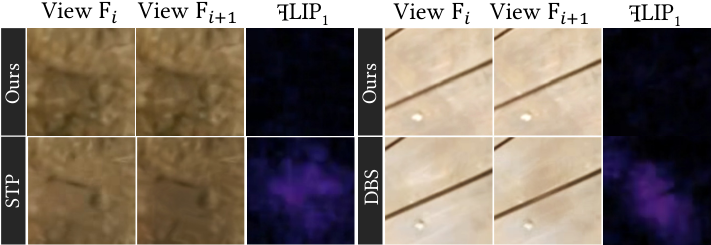}
\caption{Qualitative comparisons of ``popping'' artifacts. The zoomed-in patches are rendered from \textit{Treehill} and \textit{Garden} in the Mip-NeRF360~\cite{mip_nerf_360} dataset. View $F_{i+1}$ is warped to View $F_i$ using optical flow for alignment and comparison. The \FLIP views mark the specific popping areas. STP: StopThePop~\cite{stopthepop2024}.\label{fig:pop_cmp}}
\end{figure}

\paragraph{View Consistency. } In~\figref{fig:pop_cmp}, our sort-free rendering eliminates popping artifacts. In~\tabref{tab:fps_metric}, our method shows slightly higher $\flipview{1}$ and $\flipview{7}$ error than GES mainly because the spherical Beta function enables higher-frequency view-dependent effects. As \FLIP computes optical-flow-based warping errors and does not distinguish between view-dependent variation and actual popping, our slightly higher \FLIP values do not reflect inferior consistency.

From video results, both HTGS and SortFreeGS avoid visible popping as well. However, SortFreeGS suffers from severe occlusion leakage, while HTGS exhibits numerous floaters, both of which significantly increase \FLIP error. Additionally, MCMC-based methods tend to produce excessively large Gaussians, which frequently pop in and out between adjacent frames, producing global brightness flicker. The negative opacity strategy of SSS improves rendering quality, but introduces flickering of dark or bright patches in unseen views.

\subsection{Ablations}
\label{sec:ablation}
We conduct a series of ablation experiments. \tabref{tab:ablation} lists all quantitative metrics.


\begin{table}[t]
\centering
\caption{Ablation study. Dataset-averaged image quality and frame rate (1080p) on Mip-NeRF360 Dataset~\cite{mipnerf360} with isolated algorithm choices. Ours (base): reverting the color representation and densification strategy to the original GES settings; w/ 2 layers: using 2 surfel layers; w/o trans. grad: stopping transmittance gradients from Gaussians to surfels. \label{tab:ablation}}
\tabcolsep=0.3cm
\renewcommand\arraystretch{1.1}
\footnotesize
\begin{tabular}{l|cccc}
\hline
\multicolumn{1}{c|}{} & SSIM$\uparrow$  & PSNR$\uparrow$  & LPIPS$\downarrow$ & FPS$\uparrow$ \\ \hline
Ours (full)                  & \cellcolor{c2}0.827 & \cellcolor{c1}28.11 & \cellcolor{c1}0.196 & 472 \\
Ours (base)        & 0.821 & 27.72 & \cellcolor{c2}0.197 & \cellcolor{c2}480 \\
w/ 2 layers               & 0.810 & 27.02 & 0.223 & \cellcolor{c1}578 \\
w/ 4 layers               & \cellcolor{c1}0.830 & \cellcolor{c2}28.10 & \cellcolor{c1}0.196 & 278 \\
w/o trans. grad         & 0.825 & 27.82 & 0.206 & 477 \\
w/o $\mathcal{L}_s$             & 0.824 & 28.05 & 0.198 & 462 \\
w/o $\mathcal{L}_{scale}$         & 0.824 & 28.07 & 0.203 & 461 \\
w/o $\mathcal{L}_t$             & 0.822 & 27.86 & 0.209 & 475 \\ \hline
\end{tabular}
\end{table}

\paragraph{Reverting to GES Configurations.} To isolate the impact of our core improvements, this variant follows the original GES setup. Specifically, we replace the SB color representation with SH, and revert our MCMC-based densification (refer to~\appref{sec:training}) to the original GES strategy. In~\tabref{tab:ablation}, this variant, denoted as \textbf{Ours (base)}, still achieves an average PSNR of 27.72 on the MipNeRF360 dataset, outperforming GES and AbsGS~\cite{ye2024absgs}, and is close to SSS~\cite{zhu20253d}. This improvement comes solely from our novel representation and rendering model and is orthogonal to densification or primitive choices.

\begin{figure}[t]
\includegraphics[width=1.0\linewidth]{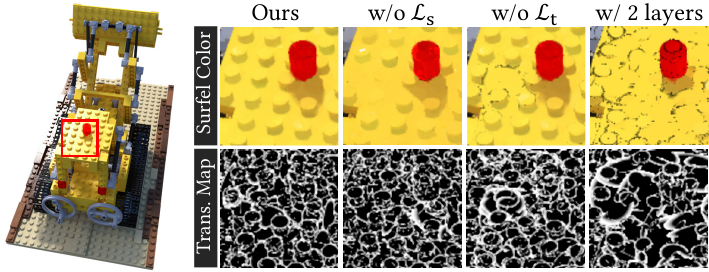}
\caption{Qualitative comparisons of rendered surfel colors without $\mathcal{L}_s$ and $\mathcal{L}_t$, and with only first 2 layers peeled. The transmittance maps $T_1^s$ show the distribution of surfels. Other ablation settings show geometry missing or background leakage (background set to black during optimization).\label{fig:abla_l3}}
\end{figure}

\paragraph{Impact of the Number of Peeling Layers.} With only the first 2 layers, any overlapping semi-transparent regions inevitably cause background color leakage, while the opaque regions in the 3rd layer can block the remaining transmittance, making it possible that the remaining transmittance reaches zero everywhere. In~\figref{fig:abla_l3}, using 3-layer peeling with $\mathcal{L}_t$ loss eliminates background leakage, whereas 2-layer peeling exhibits leakage artifacts at various locations. In~\tabref{tab:ablation}, peeling 4 layers offers negligible quality gain but incurs a significant frame-rate drop, since surfel rendering does not model appearance details and 4-layer peeling increases memory bandwidth in OpenGL by breaking 4-float-alignment packing. Thus, 3-layer peeling is our choice after balancing quality and performance.

\begin{figure}[t]
\includegraphics[width=1.0\linewidth]{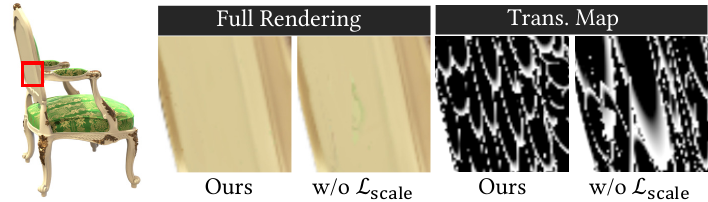}
\caption{Qualitative comparisons of images rendered without $\mathcal{L}_{scale}$. Transmittance maps $T_1^s$ show the distribution of surfels.\label{fig:abla_scale}}
\end{figure}

\begin{figure}[t]
\includegraphics[width=1.0\linewidth]{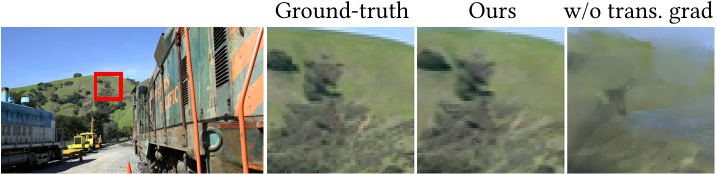}
\caption{Qualitative comparisons of images rendered with transmittance-mediated gradients stopped.\label{fig:abla_detach}}
\end{figure}

\paragraph{Impact of $\mathcal{L}_s$, $\mathcal{L}_t$ and $\mathcal{L}_{scale}$.} 
$\mathcal{L}_s$ enforces geometric alignment with ground-truth colors, preventing missing geometry as shown in~\figref{fig:abla_l3}.
Removing $\mathcal{L}_t$ increases the probability of background leakage caused by the stack of overlapping semi-transparent regions.
In~\figref{fig:abla_scale}, $\mathcal{L}_{scale}$ regularizes surfel scaling for spatial uniformity, reducing color leakage from chaotic scaling distribution. 
\paragraph{Impact of New Gradient Path.} During rendering, transmittance values link surfels and Gaussians, enabling surfel geometry parameters $S_g=\{\mathbf{p}_i, \mathbf{r}_i, \mathbf{s}_i\}_{i=1}^N$ to receive gradients not only directly from image loss $\frac{\partial\mathcal{L}}{\partial C}\frac{\partial\mathcal{C}}{\partial C_s}\frac{\partial C_s}{\partial S_g}$ but also indirectly through transmittance-mediated Gaussian gradients $\frac{\partial\mathcal{L}}{\partial C}\frac{\partial\mathcal{C}}{\partial C_G}\frac{\partial C_G}{\partial t_s}\frac{\partial t_s}{\partial S_g}$.   
In~\figref{fig:abla_detach}, stopping these gradient paths leads to blurrier details and prominent small surfels, similar to the artifacts shown in GES in~\figref{fig:ges_cmp}, confirming the importance of treating surfels and Gaussians as a fully coupled differentiable system, compared with differentiable surfels only.

\section{Limitations and Future Work}
Our work is subject to some limitations. Similar to GES, DP-GES shows artifacts in transparent or semi-transparent objects (\eg, glass windows), as our surfels are mainly opaque. Nevertheless, our improved representation produces noticeably better reconstruction even in these challenging cases (see the glass bottle in~\figref{fig:teaser}). Second, our surfel is less flexible than the fully translucent surfels in 2DGS~\cite{2dgs2024}, requiring extra initialization and relatively long training time. Moreover, while spherical Beta function enhances view-dependent effects, it also increases the risk of overfitting, as observed in DBS~\cite{Beta25Liu}.

DP-GES significantly improves the spatial distribution of surfels. In future, we plan to further explore the potential of surfels as a coarse geometry and appearance representation, and apply to high-quality mesh extraction, physically-based interactions, and shadow casting. 

\section*{Acknowledgments}
This work is partially supported by NSF China (No. 62227806, 62421003) and the XPLORER PRIZE.

{
    \small
    \bibliographystyle{ieeenat_fullname}
    \bibliography{main}
}

\clearpage
\setcounter{page}{1}
\maketitlesupplementary

In the supplementary material, \secref{sec:epsilon}–\secref{sec:appendix_geo} describe more technical details, while the remaining sections present additional experimental results.

\section{Determining Depth Margin}
\label{sec:epsilon}
In~\secref{sec:rendering}, we use the depth margin $\epsilon_s$ to prevent Gaussians intersecting with surfels from being partially truncated during the depth test. Specifically, we first compute the initial value of $\epsilon_i=\frac{5}{2}(\mathbf{s}_i^X+\mathbf{s}_i^Y)$ in the same way as GES, and then, for each surfel, we query its 16 nearest neighboring surfels and take the median of their $\epsilon$ values as the final one. Our motivation stems from the fact that using a fixed constant margin can prevent truncation but may lead to occlusion leakage in fine structures, as discussed in GES~\cite{GES25ye}. To address this, GES defines the margin based on surfel scaling, achieving geometry-adaptive control to mitigate such leakage. However, we observed that this strategy is not sufficiently robust: when a large surfel is located near fine structures composed of many small surfels, its large depth margin can still cause occlusion leakage. Our modification enforces a locally smooth distribution of $\epsilon$, further reducing the probability of such leakage.

\section{Training Details}
\label{sec:training}
\paragraph{Joint Optimization.} During this stage, we adopt the densification and pruning strategies of GES with minor modifications. Specifically, for surfels, we prune surfels that are either heavily occluded or extremely small, determined by the number of pixels they cover in the first peeled layer. The pruning thresholds are the same as GES. 

\begin{figure}[t]
\includegraphics[width=1.0\linewidth]{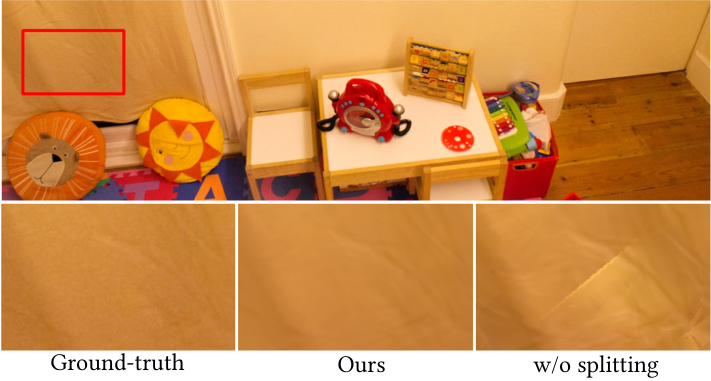}
\caption{Qualitative comparisons of images rendered without our large Gaussian splitting strategy during training.\label{fig:abla_split}}
\end{figure}

For Gaussians, we introduce an extra step to split overly large Gaussians. We found that DBS~\cite{Beta25Liu} tends to generate many excessively large Gaussians due to the added positional noise, and such Gaussians often intersect with surfels, causing partial occlusion artifacts (see~\figref{fig:abla_split}). Although our depth tolerance margin $\epsilon$ helps mitigate this issue, it remains ineffective for extremely large Gaussians. Therefore, every 2000 iterations, we query the 16 nearest surfels for each Gaussian, compute the average $\epsilon$ among them, and if a Gaussian’s average scale exceeds twice this value, we identify it as a potentially oversized Gaussian and split it into two smaller Gaussians following the same splitting strategy as in 3DGS~\cite{3DGS}. For the depth margin $\epsilon$, since the surfel scale remains stable after initialization, it is computed only at the beginning and once again at the 4k-th iteration of joint optimization. The joint optimization converges after around 25k iterations on average.

\paragraph{Initialization.} Although our surfels are differentiable within their semi-transparent boundary regions, their optimization efficiency remains limited compared to fully continuous representations. Therefore, similar to GES, we initialize the surfel positions and scales using 2DGS~\cite{2dgs2024}. Specifically, we take the sparse SfM point cloud as the initial input and train a 2DGS model for 20k iterations using the same differentiable rendering pipeline and loss functions as in the original 2DGS. The positions and scales of the final 2D Gaussians with opacity values greater than 0.8 are then used to initialize our surfels. This threshold is the same as in GES~\cite{GES25ye}. 

For Gaussians, we incorporate the MCMC-based densification from DBS~\cite{Beta25Liu} via a fast initialization step to control the primitive count and enhance reconstruction quality. Specifically, using the sparse SfM point cloud as input, we train DBS for 20k iterations with its Beta kernel replaced by a fixed Gaussian kernel, utilizing the resulting positions and scales for our Gaussian initialization. To accelerate this process, the initial number of Gaussians is kept significantly smaller than the final target. Following this MCMC phase, we continue adding Gaussians based on image-space errors during training, identically to GES, until the predefined limit is reached.

\begin{figure}[t]
\includegraphics[width=1.0\linewidth]{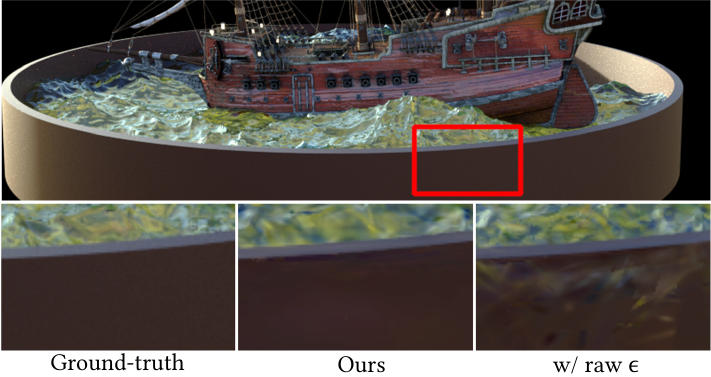}
\caption{Qualitative comparisons of images rendered with the original $\epsilon$ strategy used in GES~\cite{GES25ye}.\label{fig:abla_epsilon}}
\end{figure}

\section{OpenGL Implementation}
\label{sec:opengl}
As demonstrated in~\secref{sec:optim}, we develop an OpenGL-based renderer for real-time rendering after training, fully leveraging the efficiency of our sort-free rendering scheme within the traditional graphics pipeline. The OpenGL-based renderer also consists of two main passes: 

In the first pass, we perform depth peeling for the nearest three surfel layers. During each rasterization step, we write not only the depth but also the surfel ID into the buffer. This additional ID information enables us to correctly identify and handle interleaved surfels, preventing the issue where two intersecting surfels could otherwise be peeled only once along their intersection line. After the three peeling passes, a post-processing shader blends the peeled layers to generate the final surfel color map, while also writing the $\epsilon$-adjusted depth into the depth buffer. The peeled depth maps $\{D_i^s\}_{i=1}^2$ and transmittance maps $\{T_i^s\}_{i=1}^2$ are packed into a four-channel texture. Since Gaussians beyond $D_3^s$ are culled by the hardware depth test, storing $D_3^s$ and $T_3^s$ is unnecessary. 

In the second pass, Gaussians are splatted onto the screen with depth testing enabled and depth writing disabled. During this process, the shader queries the peeled depth and transmittance values from the four-channel texture to accumulate color and weight contributions. Finally, another post-processing shader performs weight normalization, producing the final rendering result. Our method is entirely based on the programmable graphics pipeline and does not rely on compute shaders, making it easy to be integrated into existing rendering engines and mobile devices.

\begin{figure*}[t]
\includegraphics[width=1.0\linewidth]{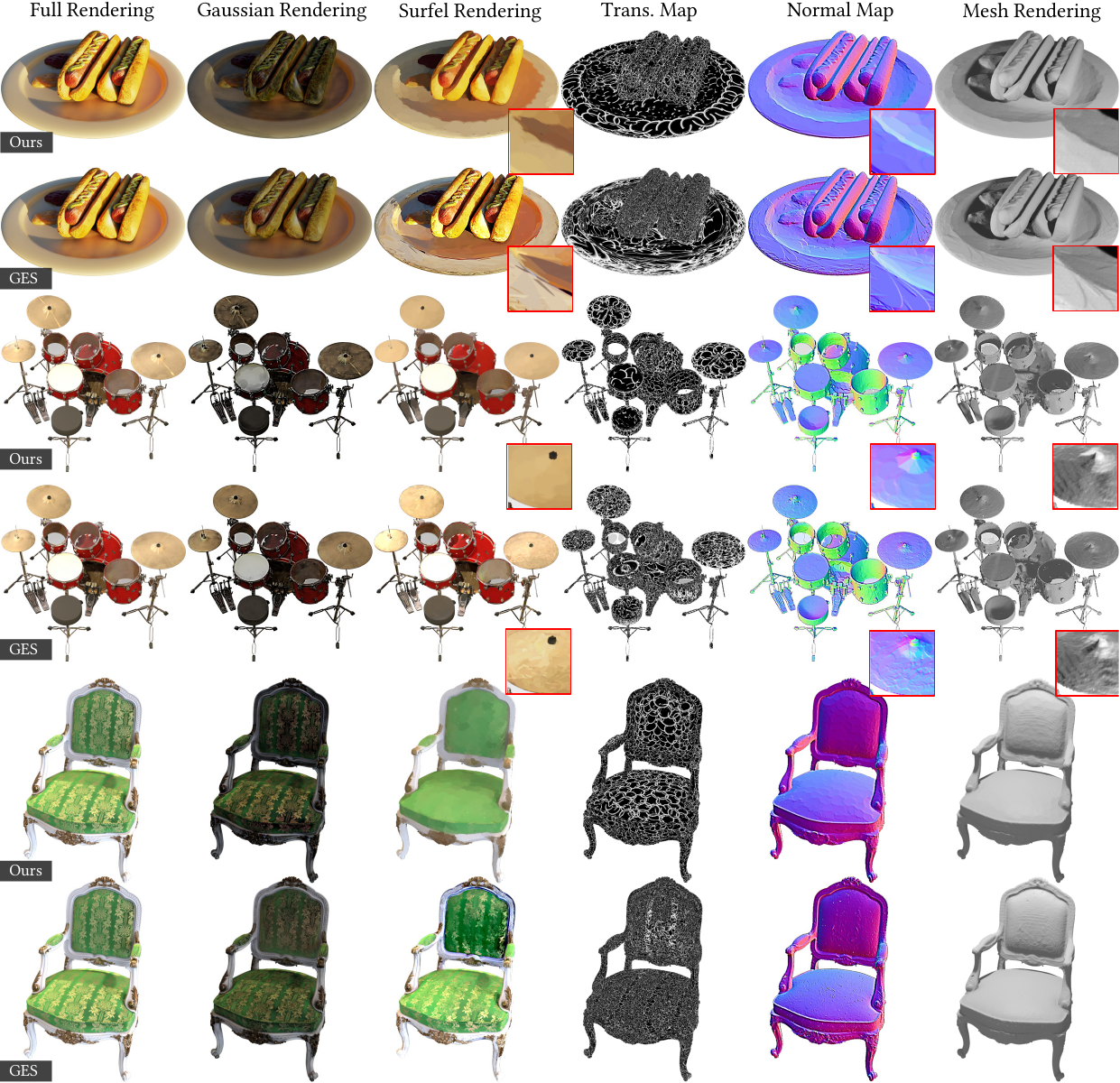}
\caption{Qualitative comparisons between our DP-GES and GES~\cite{GES25ye} are shown across three aspects: surfel-Gaussian decomposition rendering, surfel distribution, and surfel geometry. For decomposition rendering, we visualize the Gaussian and surfel components as $C_G/(W_s+W_G)$ and $C_s/(W_s+W_G)$, respectively. Surfel distribution is illustrated by the transmittance map $T_1^s$ in DP-GES, with GES processed in a similar way. Surfel geometry is represented by the normal map computed from depth and the mesh extracted from surfels. From top to bottom: scenes are \textit{Hotdog}, \textit{Drums} and \textit{Chair} from the NeRF Synthetic Dataset~\cite{nerf}.\label{fig:ges_geo_dec}}
\end{figure*}

\section{Geometry Regularization}
\label{sec:appendix_geo}
To enable the surfel representation to better capture the scene geometry, which is essential for downstream tasks such as mesh extraction, we optionally adopt a geometry regularization loss $\mathcal{L}_{geo}=\lambda_4\mathcal{L}_{sd}+\lambda_5\mathcal{L}_{sn}+\lambda_6\mathcal{L}_{sdn}$, where $\lambda_4=0.1$, $\lambda_5=0.05$ and $\lambda_6=0.05$, respectively.
\begin{equation}
    \mathcal{L}_{sd}=\mathcal{L}_1(D_{supv}, D_s),
\end{equation}
\begin{equation}
    \mathcal{L}_{sn}=\frac{1}{HW}\sum_{\hat{\mathbf{x}}}(1-N_{supv}(\hat{\mathbf{x}})\cdot N_s(\hat{\mathbf{x}})),
\end{equation}
\begin{equation}
    \mathcal{L}_{sdn}=\frac{1}{HW}\sum_{\hat{\mathbf{x}}}(1-N_{supv}(\hat{\mathbf{x}})\cdot \nabla(D_s(\hat{\mathbf{x}}))),
\end{equation}
where $\nabla(\cdot )$ is the depth-to-normal operator, the reference depth maps $D_{supv}$ and normal maps $N_{supv}$ are generated in the last iteration of 2DGS initialization, and the depth map $D_s$ and normal map $N_s$ of surfels are rendered by alpha blending, following the same procedure as~\equref{equ:s_render}.
\begin{table}[t]
\centering
\caption{Ablation study. Dataset-averaged image quality and frame rate (1080p) on Mip-NeRF360 Dataset~\cite{mipnerf360}. Base/Full: using our base/full model; w/o splitting: without the large-Gaussian splitting strategy; w/ raw $\epsilon$: using the original $\epsilon$ computation from GES~\cite{GES25ye}. \label{tab:ablation_appendix}}
\tabcolsep=0.19cm
\renewcommand\arraystretch{1.1}
\small
\begin{tabular}{l|cccc}
\hline
\multicolumn{1}{c|}{} & SSIM$\uparrow$  & PSNR$\uparrow$  & LPIPS$\downarrow$ & FPS$\uparrow$ \\ \hline

GES                   & 0.813 & 27.38 & 0.208 & 675 \\
Base, w/o trans. grad & 0.818 & 27.57 & 0.201 & 482 \\
Base                  & 0.821 & 27.72 & 0.197 & 480  \\
Base, w/ MCMC dens.   & 0.826 & 27.98 & 0.195 & 470  \\
Full                 & 0.827 & 28.11 & 0.196 & 472 \\
\hline
Full, w/o splitting        & 0.825 & 28.04 & 0.197 & 479 \\
Full, w/ raw $\epsilon$        & 0.827 & 28.08 & 0.199 & 469 \\ \hline
\end{tabular}
\end{table}
\section{Additional Comparisons with GES}
\paragraph{Surfel-Gaussian Decomposition Rendering.}
In the leftmost 3 columns of~\figref{fig:ges_geo_dec}, we visualize the surfel-Gaussian decomposition rendering compared with GES. In \textit{Chair}, GES preserves high-frequency texture in surfel colors because its image-loss-based surfel initialization inevitably embeds appearance details, deviating from the intended coarse-geometry representation. In contrast, DP-GES naturally transfers high-frequency details from surfels to Gaussians during joint training, resulting in smoother surfel rendering without visual noise (see the zoomed-in patches of \textit{Hotdog} and \textit{Drums}).

\paragraph{Surfel Distribution and Geometry Reconstruction.}
In this experiment, we compare our DP-GES with GES, with geometry regularization enabled for both methods. In the rightmost 3 columns of~\figref{fig:ges_geo_dec}, the joint optimization in DP-GES keeps surfels well-regularized, yielding a more uniform and coherent distribution than GES. This prevents geometry and appearance artifacts: GES often produces needle-like surfels that manifest as sharp spikes, resulting in non-smooth surfaces in the extracted mesh.

\section{Additional Ablations}
We present quantitative additive ablations and additional results in~\tabref{tab:ablation_appendix}. The additive study demonstrates a clear and continuous improvement as we progressively upgrade the original GES to our full model. Applying our rendering model to GES while disabling the new gradient path (\textit{Base, w/o trans. grad}) alleviates the aliasing issue, yet the reconstruction remains suboptimal. Enabling the gradient path (\textit{Base}) enhances the reconstruction quality. Incorporating the MCMC-based densification strategy from DBS~\cite{Beta25Liu} (\textit{Base, w/o MCMC dens.}) yields additional gains, and replacing the SH color with SB color representation (\textit{Full}) brings a marginal but consistent quality increase. These steady improvements validate that our core contributions are highly effective and orthogonal to existing auxiliary techniques.

\paragraph{Impact of Improved Depth Margin Strategy.} In~\figref{fig:abla_epsilon}, smooth structures (\eg, pot rims) and detailed structures (\eg, waves) are typically fitted by large and small surfels, respectively. Using the same $\epsilon$ strategy as in GES, occlusion leakage occurs in regions where large and small surfels are adjacent. Our $\epsilon$ computation incorporates the sizes of neighboring surfels, yielding smoother local transitions and reducing such leakage.

\paragraph{Impact of Large Gaussian Splitting.}
In~\figref{fig:abla_split}, our Gaussian splitting strategy mitigates the issue of excessively large Gaussians produced during DBS initialization. Without splitting, these large Gaussians can extend across surfels and be abruptly truncated by the depth test, producing visual artifacts. By splitting them into smaller Gaussians, we reduce the probability of such depth-test cutoffs.

\end{document}